\documentclass[twocolumn]{aastex631}

\newcommand{\msun}{\mbox{M}_\odot}
\newcommand{\mearth}{\mbox{M}_\oplus}
\newcommand{\pab}{\mbox{Pa}\beta}
\newcommand{\brg}{\mbox{Br}\gamma}
\newcommand{\lacc}{L_{\rm acc}}
\newcommand{\macc}{\dot{M}_{\rm acc}}
\newcommand{\lstar}{L_\star}
\newcommand{\mstar}{M_\star}
\newcommand{\rstar}{R_\star}

\newcommand{\lbol}{L_{\rm bol}}
\newcommand{\mdisk}{M_{\rm disk}}

\usepackage{multirow}
\usepackage{rotating}
\usepackage[graphicx]{realboxes}

\begin{document}

\title{The relation between the Mass Accretion Rate and the Disk Mass in Class I Protostars}

\author[0000-0002-5261-6216]{Eleonora Fiorellino}
  \affiliation{Konkoly Observatory, Research Centre for Astronomy and Earth Sciences, E\"otv\"os Lor\'and Research Network (ELKH), Konkoly-Thege Mikl\'os \'ut 15-17, 1121 Budapest, Hungary}
  \affiliation{CSFK, MTA Centre of Excellence, Budapest, Konkoly Thege Mikl\'os \'ut 15-17., H-1121, Hungary}
  \affiliation{INAF-Osservatorio Astronomico di Capodimonte, via Moiariello 16, 80131 Napoli, Italy}
\author[0000-0002-9470-2358]{{\L}ukasz Tychoniec}
  \affiliation{European Southern Observatory, Karl-Schwarzschild-Strasse 2, 85748 Garching bei München, Germany}

\author[0000-0003-3562-262X]{Carlo F. Manara}
  \affiliation{European Southern Observatory, Karl-Schwarzschild-Strasse 2, 85748 Garching bei München, Germany}

\author[0000-0003-4853-5736]{Giovanni Rosotti}
\affiliation{Leiden Observatory, Leiden University, PO Box 9513, NL-2300 RA Leiden, the Netherlands}
\affiliation{School of Physics and Astronomy, University of Leicester, University Road, Leicester LE1 7RH, UK}

\author[0000-0002-0666-3847]{Simone Antoniucci}
\affiliation{INAF-Osservatorio Astronomico di Roma, via di Frascati 33, 00078, Monte Porzio Catone, Italy}

\author[0000-0002-4283-2185]{Fernando Cruz-S\'aenz de Miera }
\affiliation{Konkoly Observatory, Research Centre for Astronomy and Earth Sciences, E\"otv\"os Lor\'and Research Network (ELKH), Konkoly-Thege Mikl\'os \'ut 15-17, 1121 Budapest, Hungary}
  \affiliation{CSFK, MTA Centre of Excellence, Budapest, Konkoly Thege Mikl\'os \'ut 15-17., H-1121, Hungary}
  
\author[0000-0001-7157-6275]{\'Agnes K\'osp\'al}
  \affiliation{Konkoly Observatory, Research Centre for Astronomy and Earth Sciences, E\"otv\"os Lor\'and Research Network (ELKH), Konkoly-Thege Mikl\'os \'ut 15-17, 1121 Budapest, Hungary}
  \affiliation{CSFK, MTA Centre of Excellence, Budapest, Konkoly Thege Mikl\'os \'ut 15-17., H-1121, Hungary}
\affiliation{Max Planck Institute for Astronomy, K\"onigstuhl 17, 69117 Heidelberg, Germany}
\affiliation{ELTE E\"otv\"os Lor\'and University, Institute of Physics, P\'azm\'any P\'eter s\'et\'any 1/A, 1117 Budapest, Hungary}

\author[0000-0002-9190-0113]{Brunella Nisini}
\affiliation{INAF-Osservatorio Astronomico di Roma, via di Frascati 33, 00078, Monte Porzio Catone, Italy}

\begin{abstract}

The evidence of a relation between the mass accretion rate and the disk mass is established for young, Class\,II pre-main sequence stars. This observational result opened an avenue to test theoretical models and constrain the initial conditions of the disk formation, fundamental in the understanding of the emergence of planetary systems.
However, it is becoming clear that the planet formation starts even before the Class\,II stage, in disks around Class\,0 and I protostars.
We show for the first time evidence for a correlation between the mass accretion rate and the disk mass for a large sample of Class\,I young stars located in nearby ($< 500$\,pc) star-forming regions. 
We fit our sample, finding that the Class\,I objects relation has a slope flatter than Class\,II stars, and have higher mass accretion rates and disk masses. The results are put in context of the disk evolution models.


\end{abstract}


\section{Introduction} \label{sec:intro}

Young stellar objects (YSOs) evolve as a result of a complex interplay between the forming star, the circumstellar disk, where planet formation occurs, and the envelope. 
Modeling efforts to describe such intricate interplay span all ranges of the protostellar lifetime from pre-stellar cores to non-accreting young stars. However, the comparison of theoretical predictions with observations of stellar properties is limited to the well-characterized Class\,II disks of the pre-main-sequence phase \citep[][and references therein]{manara2022pp7}.

Observations of the Class\,0/I stages are especially important in the context of constraining the initial conditions for the models of disk evolution. 
In particular, a fundamental parameter which describes part of this interplay is the mass accretion rate ($\macc$) which correlates with the disk mass ($\mdisk$).
This correlation was predicted by viscous model \citep[][]{har98}, and recently confirmed by observations  \citep[][]{man16letter} for Class\,II objects. 

Since the advent of submillimeter interferometry and complete surveys of planet-forming disks, our knowledge of the disks around the youngest protostars has greatly expanded \citep{Sheehan.Eisner2017,wil19,Maury.Andre.ea2019,tyc20,tob20, miotello2022pp7}. 
At the same time, thanks to new infrared (IR) facilities, some efforts to characterize stellar properties of the youngest stars \cite[e.g.,][]{laos21,fio21} as well as available archival observations \citep{muz98, whi04, dop05, con10} show a promising way to investigate $\macc$ in the protostellar phase. 

In this letter we put recent observations of Class\,I protostars in the context of disk evolution models, both in the viscous and disk wind paradigms \citep{lodato17,Tabone.Rosotti.ea2021}, and models of early stages of core collapse and disk formation \citep{Zhao.Tomida.ea2020,Hennebelle.Commercon.ea2020}. 
We present and discuss the $\macc$\,vs.\,$\mdisk$ relation for the first time for Class\,I protostars.

\section{Sample \& Method} \label{sec:data}

This work is based on the already existing observations of protostellar sources.
The sample analysed here is composed of 26 Class\,I young stars (whose spectral index between 2 and 24\,$\mu$m is $\alpha > -0.3$, i.e. we include also Flat spectrum objects) located within 500\,pc of the Sun. 
The choice of these sources is driven by the need of an homogeneous computation of the mass accretion rate and the disk mass.  
Therefore, we collect sources whose accretion analysis is based on near-IR (NIR) spectroscopic tracers, and for which millimeter archival data from which we calculate the disk dust mass are available. These criteria were satisfied by the 3 objects from \citet[][]{nis05accretion} in Corona\,Australis cloud; the 3 objects from \citet{ant08} all within 450\,pc; 6 objects out of the 10 analysed by \citet{fio21} in NGC\,1333 cluster in the Perseus star-forming region; and 14 objects from Fiorellino et al. (submitted) out of the 40 protostars analysed therein.   
The list of targets included in the analysis is reported in Appendix\,\ref{app:sample}.

\subsection{The Mass Accretion Rate}
The mass accretion rate for all these sources was computed by using similar methods that provide comparable results. The main common assumption is that the accretion during the Class\,I stage can be described through the magnetospheric accretion scenario \citep[{for a recent review, see}][]{har16} and computed with the related equation: 
\begin{equation}
    \macc \sim \left( 1 - \frac{\rstar}{R_{in}} \right)^{-1} \frac{\lacc \rstar}{G \mstar}
\end{equation}
\noindent where $R_{in}$ is the inner-disk radius which we assume to be $R_{in} \sim 5 \rstar$ \citep{har98}, and $G$ is the gravitational constant.
\citet[][]{nis05accretion} computed the accretion luminosity by the difference between the bolometric and the stellar luminosity. 
They found that their results for Class\,I protostars were in agreement with $\lacc$ computed using empirical relations which link the HI emission lines, $\pab$ and $\brg$, with the accretion luminosity in Class\,II PMS stars from \citet[][]{muz98}. 
\citet[][]{ant08} derived the accretion rates using a self-consistent method based on the aforementioned empirical relations, the assumption that the bolometric luminosity is the sum of the accretion and stellar luminosity ($\lbol = \lacc + \lstar$), the equation of the bolometric magnitude in  $K$\,band: M$_{\rm bol} = \mbox{BC}_K + m_K + 2.5 \log(1 + r_K) - A_K - 5 \log(d/10 \mbox{pc})$, and the assumption that these sources lie on the
birthline \citep[as described by][]{Palla.Stahler1990}.
Later on, \citet[][]{fio21} and Fiorellino et al. ({\it submit.}) adopted the same self-consistent method by using the most recent empirical relations by \citet[][]{alc17} and assuming the age of these sources is between the birthline and 1\,Myr, based on {\it Spitzer}-based lifetime estimates for Class\,I and Flat objects \citep[][]{eno09,dun14}. For a detailed description of the self-consistent method we refer the reader to \citet[][]{ant08} and \citet{fio21}. 
Average errors on the mass accretion rate is 0.8\,dex \citep[][]{fio21}. 
However, we note that since young stars variability, a further 0.5\,mag uncertainty in flux \citep[][]{lorenzetti2013} should be considered. This propagates a variation on the flux of about 50\%, enlarging the uncertainties on the $\macc$ as a consequence.

We would like to focus the attention of the reader on the following observational limit. This kind of mass accretion analysis is possible only for sources where the IR veiling due to the disk and envelope is sufficiently low that we can see the photosphere. 
Usually, according to the current correspondence between Classes based on the SED spectral index and evolutionary path, the less embedded objects are the more evolved ones. 
Therefore, we can consider this sample of Class\,I representative of the brightest and more evolved Class\,I protostars. 

\subsection{The Disk Dust Mass}
For the overall sample we performed a coordinate and sources name search across the literature and also looked for archival interferometric data. 
We included a dust mass measurement in our analysis if the flux measurement was available at $<$1$"$ resolution. 
In the sub-arcsecond regime with size of the beam comparable to the disk size, the envelope contribution is usually negligible, especially for Class\,I systems where the envelope is largely dissipIn the sub-arcsecond regime with size of the beam comparable to the disk size, the envelope contribution is usually negligible, especially for Class\,I systems where the envelope is largely dissipated \citep{tyc20}. If there was no flux reported in the literature but data was available in the archive, we performed 2D Gaussian fit to the continuum image to extract the flux density.

ated \citep{tyc20}. If there was no flux reported in the literature and data was available in the archive, we performed 2D Gaussian fit to the continuum image to extract the flux density. This was done on the pipeline processed products in the archive, without any additional processing.

From the flux density ($F_\nu$) we calculated the dust mass by inverting the modified black-body equation:
\begin{equation}
\label{eq:dustmass}
M_{\rm dust}=\frac{d^2F_\nu}{{\kappa}_\nu(\beta) B_\nu(T_{\rm dust})}\,
,\end{equation}

\noindent where $d$ is the distance to the source, $B_\nu$ is the Planck function for the dust temperature $T_{\rm dust}$, and $\kappa_\nu$ is the dust opacity at the frequency of the observation $\nu$. 
The disk mass ($\mdisk$) is obtained from the dust mass assuming a typical dust-to-gas ratio of 1:100.
Eq.\,\ref{eq:dustmass} is accurate for optically thin emission, otherwise it provides a lower limit on the dust mass measurement. 
An isothermal disk with $T=30\,K$ is assumed, which is the temperature typically used for young, embedded disks. If the disks are colder, similar to the Class II systems, this would result in the increase of the total dust mass.
Dust opacity value at 1.3 mm is 0.00899\,g\,cm$^{-2}$ from \citet{Ossenkopf.Henning1994} and for observations at different wavelengths the spectral emissivity is scaled with $\beta=$\,1, which assumes some degree of grain growth \citep[][]{natta_review05}.
With uniform assumptions on dust properties we are not introducing additional discrepancy between the disks measured within different observing projects. 
The accuracy of the disk mass estimation is a matter of ongoing debate \citep[see][and references therein]{miotello2022pp7, manara2022pp7}. 
Several studies point to severe underestimation of the disk mass due to optical thickness or dust scattering \citep{Zhu.Zhang.ea2019}. 
Recent work of \cite{Sheehan.Tobin.ea2022} shows that Class 0/I disk dust masses can be overestimated -- especially on the low-mass end -- if the simplistic assumption of isothermal disk is used. 
Combined, these effects would result in increased spread of the disk masses (i.e., more massive disks would be even more massive while the low-mass end would have even lower masses).

\begin{figure}[t]
    \centering
    \includegraphics[width=0.45\textwidth]{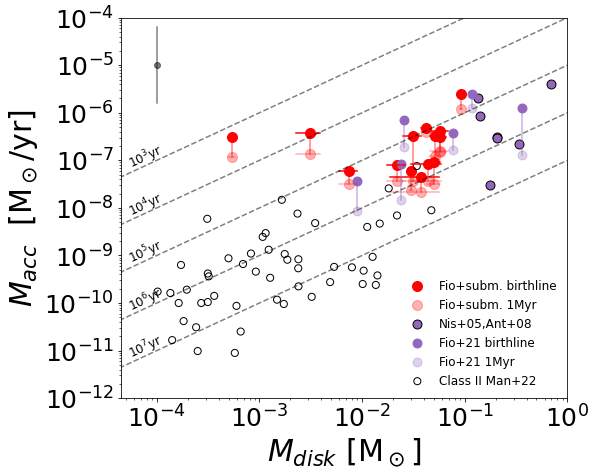}
    
\caption{Mass accretion rate vs. disk dust mass. 
    Red filled circles are the Class\,I sources where $\macc$ is the mean value between results assuming the birthline and 1\,Myr old evolutive track. Purple filled circles are other Class\,I from the literature.
    Empty circles are Class\,II from Lupus sample. Black dashed lines correspond to several disk depletion times $t_\nu = \macc/\mdisk$, from $10^3$\,yr (top) to $10^7$\,yr. A representative error bar for the mass accretion rate is shown in grey on the top left corner of the plot.}
    \label{fig:histo_disk}
\end{figure}

\section{results and discussion}
\label{sec:results}
Viscous models of disk evolution predict a strong correlation between $\macc$ and $\mdisk$ \citep[e.g.][]{har98, dullemond06, rosotti17}. 
In the last years, this relation has been investigated in CTTSs samples thanks to accurate measurements of the mass accretion rate and disk mass for several nearby ($< 500$\,pc) star forming regions \citep[see][and references therein]{manara2022pp7}.
In brief, these works show that the predicted trend is confirmed by observations with a spread of about $\sim 1$\,dex \citep{man16letter}.
Moreover, the spread is still present for old star-forming regions as Upper Scorpius \citep{man20}. This is contrary to expectations of the viscous model which predicts a decrease of the spread with the age of the CTTSs population.
An interesting missing piece of information in this debate is whether the strong correlation is still valid in the earlier stages, where the viscous timescale starts to be comparable or larger than the YSOs lifetime itself. In particular, with information of the accretion rates and disk masses in the earlier stages of evolution we can constrain initial conditions for the disk evolution models.

\subsection{$\macc/\mdisk$ in Class\,I protostars} 
Fig.\,\ref{fig:histo_disk} shows the mass accretion rate as a function of the total disk mass for our sample of Class\,I protostar.
For each source, we plot the $\macc$ derived assuming the sources on the birthline and the one derived assuming they are 1\,Myr old, which correspond to the edges of the possible values. 
We also plot Class II disks in Lupus, for which properties were obtained by \citet{manara2022pp7}, for a comparison of our sample with more evolved sources.
The depletion time $t_\nu = \macc / \mdisk$ from $10^3$\,yr to $10^7$\,yr are plotted (black dashed lines).

The plot shows that the more disk massive Class\,I systems are matching a trend seen for Class\,II disks in Lupus star-forming region  from \citet{man16letter} with depletion times $10^4\,{\rm yr} < t_\nu < 10^7\,{\rm yr}$. An exception is represented by two protostars, the ones with the less massive disk, and the only two sources laying on a region of the $\macc \, vs. \, \mdisk$ described by depletion times shorter than $10^4$\,yr . 
Considering the overall sample of Class\,I, there is a large scatter on $\macc \, vs \, \mdisk$ distribution. This trend is particularly notable in sources with $\mdisk < 10^{-2}\,\msun$. 


To investigate the relation between Class\,I and II YSOs samples, we performed a two-sample Kolmogorov–Smirnov test (KS test) to quantify the difference of their $\mdisk/\macc/{\rm yr}$ distributions. 
We obtained that the probability that the Class\,I and II samples could have been drawn from the same probability distribution is 0.2 considering Class\,I on the birthline, and 0.4 considering Class\,I being 1\,Myr old.
Assuming the two ages as ``limits" for our sample, the probability that our sample of Class\,I YSOs is drawn from the same probability distribution of the Lupus Class\,II is $0.2 < p < 0.4$. 
We note that by assuming 1\,Myr as the Class\,I sample age, the probability to have the same statistical distribution of Class\,II is not negligible.
This result shows that the $\mdisk/\macc/{\rm yr}$ distribution can be separated depending on the evolutionary stage of the disk, evolving with the age.    
This can be due to a different evolution of the disk during the Class\,I stage, when the refuel from the envelope is not negligible.
Fig.\,\ref{fig:histo_disk} also suggests that Class\,I disks accrete more material on the central star than Class\,II.

\begin{figure}
    \centering
    \includegraphics[width=0.48\textwidth]{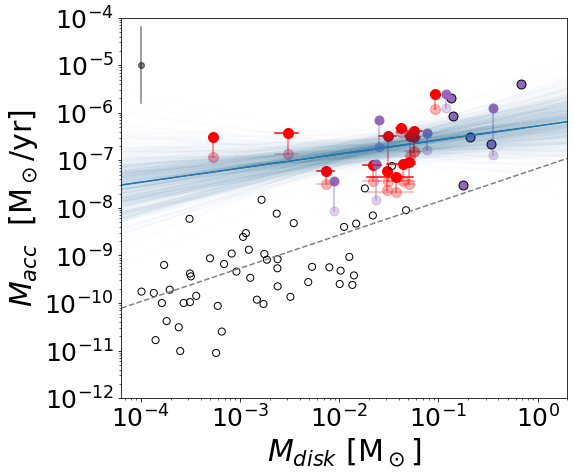}
    \caption{Mass accretion rate vs. disk dust mass. 
    Symbols are as in Fig.\,\ref{fig:histo_disk}. The blue line corresponds to our best fit for the overall sample of Class\,I YSOs, while the light blue lines are a subsample of the results of some chains. The dashed grey shows the best fit for Class\,II YSOs} \citep[][]{man16letter}.
    \label{fig:mdisk_macc}
\end{figure} 

In Fig.\,\ref{fig:mdisk_macc} we performed a linear regression fit of the Class\,I protostars sample (light-blue dashed line) considering the mean value of the $\macc$ between the one computed assuming sources on the birthline and 1\,Myr old, having as error the standard deviation plus the intrinsic uncertainty. We used the hierarchical Bayesian method by \citet[][]{kel07} which considers error in both the axes of the plot. 
We found the following relation $\log (\macc) = (0.3 \pm 0.2) \log (\mdisk) + (-6.3 \pm 0.3)$ with a standard deviation of $0.4 \pm 0.1$.  
We compare our fit with the one performed for Class\,II \citep[][]{man16letter}. 
The grey dashed line in Fig.\,\ref{fig:mdisk_macc} shows the Class\,II slope of $0.7 \pm 0.2$ moved upwards in the plot, at the same intercept we obtained.
Our best fit shows a slope flatter than the one obtained for Class\,II,  lying always above the Class\,II sample.
We note that if we remove from the fitted sample the two sources with the smallest disk mass, we find a slope of $1.1 \pm 0.2$, compatible within the error with the Class\,II slope. This could suggest that to determine how and how much the Class\,I and II YSOs $\macc - \mdisk$ distributions are different, we should analyse Class\,I objects with low disk mass, comparable with the disk mass typical of Class\,II sources, to verify whether these two sources are outliers or Class\,I show a flatter slope in general.
On the other edge of the distribution, we expect that more embedded (and younger) protostars, with $M_{dust} > 20\, \mearth$ (i.e. $\mdisk > 6 \times 10^{-4} \msun$), would lay above the current distribution, i.e. higher $\macc$ and similar $\mdisk$, as suggested by simulations by \citet[][]{Hennebelle.Commercon.ea2020} which provide almost constant value of about $1.5 \times 10^{-2}$\,$\msun$ for protostellar disk mass until 0.16\,Myr.
In this case, it would be possible to exclude the same evolutionary path (same slope) with different initial disk conditions.

Limiting our discussion to the Class\,I we analysed, we can state that they tend to have higher mass accretion rate when contrasted with Class\,II disks with a comparable disk mass, and that the $\macc/\mdisk$ distribution is flatter in Class\,I than in Class\,II YSOs. 
This implies that, assuming $\mdisk$ and $\macc$ constant within the Class\,I lifetime \citep[$\sim 0.54$\,yr,][]{eno09,dun14}, the disk should be dissipated within $10^5-10^6$\,yr for most of the protostars in our sample. 
But we know that the presence of the disk on the YSOs systems last until $\sim 10^{6}-10^7$\,yr.
Possible solutions to this discrepancy are:
\newline ({\it a}) $\macc$ is not constant. For example, it can be possible that during this so short evolutionary stage, $\macc$ decreases rapidly, reaching lower values typical of Class\,II objects, not totally dissipating the disk; 
\newline ({\it b}) $\macc$ is constant during the Class\,I protostellar stage, but the disk is fuel by an ``extra mass" coming from the envelope.

\begin{figure}
    \centering
    \includegraphics[width=0.48\textwidth]{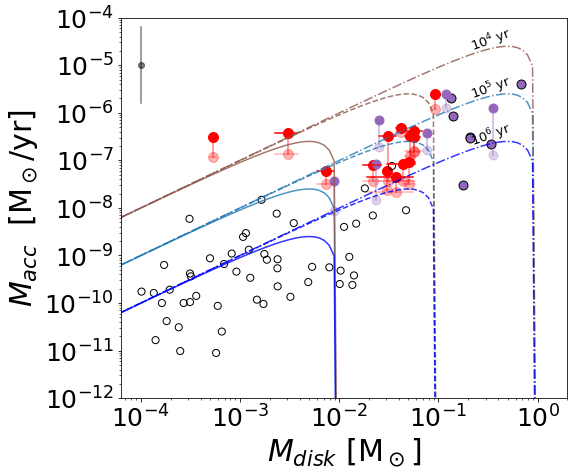}
    \caption{Mass accretion rate vs. disk dust mass. Symbols are as in Fig.\,\ref{fig:histo_disk}.
    Brown, light blue, and blue dashed lines are isochrones from \citet[][]{lodato17} at $t_\nu = 10^4$, 10$^5$, and 10$^6$ yr, respectively. For each $t_\nu$ we show in solid, dashed, and dash-dotted lines isochrones for three different initial disk masses (0.01, 0.1, and $1 \msun$, respectively).}
    \label{fig:mdisk_macc_ages}
\end{figure} 

While we still have to collect suitable data to check predictions on the $\macc/\mdisk$ distribution which consider a continuous fuel of material from the envelope to the disk \citep[i.e.][]{Hueso2005}, the scenario in which $\macc$ decreases rapidly to conserve the disk is described \citet[][]{lodato17}. 
To investigate this hypothesis, we plot on Fig.\,\ref{fig:mdisk_macc_ages} the comparison between our results and isochrones by \citet[][]{lodato17} for  $\mdisk/\macc = 10^{4}$, $10^{5}$, and $10^{5}$\,yr, and initial disk mass of 0.01, 0.1, and 1\,$\msun$. 
Results suggest that there is no correlation between the assumed age of the stars and the viscous timescales of the isochrones, 
and that our data can be reproduced by a variety of isochrones with different $t_\nu$ and initial disks conditions.
We note that most of the sources cluster in the region between $10^5$ and 10$^6$\,yr and $M_{disk, 0}$ between 0.01 and 1\,$\msun$. 

\subsection{The $\macc/\mdisk$ Evolution} 

The comparison between Class\,I and Class\,II $\macc/\mdisk$ distributions leads to the question: are our Class\,I sources going to reproduce Lupus Class\,II population in 1\,Myr?
A positive answer to this question would imply not only that the disk is able to survive until Class\,II stage, but also that the $\macc$ and $\mdisk$ values are compatible with the Class\,II population ones.

\begin{figure}
    \centering
    \includegraphics[width=0.48\textwidth]{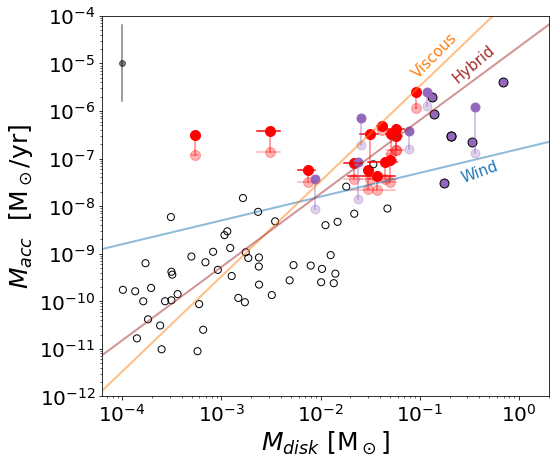}
    \caption{Mass accretion rate vs. disk dust mass. Symbols are as in Fig.\,\ref{fig:histo_disk}.
    Solid lines represent evolutionary tracks assuming different models: purely viscous (orange), hybrid of viscous and MHD wind (brown), and dominated by MHD wind (light blue).}
    \label{fig:mdisk_macc_ages1}
\end{figure}

In order to put the measured accretion rates and disk masses in the context of disk evolution models, we plot in Fig.\,\ref{fig:mdisk_macc_ages1}  evolutionary tracks of different evolution models, namely viscous evolution, pure disk wind evolution, and hybrid case in which both modes of evolution are in place. Each of the models is briefly described below.

According to \citet[][]{lodato17}, the evolutionary track of a viscously evolving disk is prescribed as : 

\begin{equation}
\label{eq:viscous_model}
 \macc =  \frac{1}{2 (2 - \gamma)} \frac{M_0}{t_{\nu}} \left(\frac{M_{\rm disk}}{M_0}\right) ^{5 - 2\gamma}
\end{equation}

\noindent where $M_0$ is the initial disk mass, t$_\nu$ is the aforementioned viscous timescale, and $\gamma$ is a factor depending on the value of the viscosity and the radius of the source. For $\gamma$ we assume 1.5, following value assumed in \cite{lodato17}, where 1.5 is showed to be consistent with Minimum-Mass Solar Nebula and the value between 1.2 and 2 is needed to reproduce the evolution of disks in Lupus. We set $M_0$=1 $\msun$ to match presented samples.

Evolution of the disk under both MHD disk wind and viscous effects (i.e. hybrid model) follows a relation described in \cite{Tabone.Rosotti.ea2021}:  
\begin{equation}
 \macc =  \dot{M}_{\rm acc,0}  \left(\frac{M_{\rm disk}}{M_0}\right) ^\frac{\psi + 3 + 4\xi}{\psi + 1 + 2\xi},
\end{equation}

\noindent where  $\dot{M}_{\rm acc,0}$ is the initial accretion rate, $\psi$ is the ratio between wind and turbulent torque, $\xi$ is the mass ejection index, which is a derivative of ln($\macc$) over ln(r) as defined in \cite[][]{Ferreira.Pelletier1995}. 
Values of the parameters used to compare the model with our results on Fig. \ref{fig:mdisk_macc_ages1} are: $\psi$=3, $\xi=0.1875$ following prescription in \cite{Tabone.Rosotti.ea2021} and $\dot{M}_{\rm acc,0}=10^5$ yr is used to best fit the data.
In a disk where accretion and mass-loss evolution is purely dependent on the MHD wind, the relation can be parametrized as follows \citep[][]{Tabone.Rosotti.ea2021}.\\
\begin{equation}
 \macc =  \dot{M}_{\rm acc,0}  \left(\frac{M_{\rm disk}}{M_0}\right)^{1-\omega},
\end{equation}

\noindent where $\omega$ is an index that describes the dependence of the disk wind torque on characteristic surface density. 
For the representative comparison we select intermediate value of $\omega$=0.5 and $\dot{M}_{\rm acc,0}=10^5$yr to best fit the data. \citet{Tabone.Rosotti.ea2021} showed that the slope of the evolutionary track changes when MHD\,disk wind effect are introduced. The slope becomes shallower for disk wind with magnetic field strength decreasing with time, up to becoming completely flat (constant accretion rate with decreasing mass) for constant magnetic field strength with disk evolution. 
The pure viscous model prescribed by Eq.\,\ref{eq:viscous_model} is shown in Fig.\, \ref{fig:mdisk_macc_ages1} (orange line), together with the evolutionary tracks of an hybrid model of viscous and MHD disk wind (brown line), and another hybrid model where MHD disk wind largely dominates the evolution (brown dashed line), reproduced using Eqs.\,60 and 61 in \cite{Tabone.Rosotti.ea2021}, respectively.
Fig.\,\ref{fig:mdisk_macc_ages1} suggests that we can conclude that the pure viscous model does not describe well the transition between our Class\,I sample and Lupus Class\,II, at least, not for the protostars with disk masses $\ll 10^{-2}\,\msun$ and $> 0.1$\,$\msun$.
Also the purely MHD disk wind evolution of the accretion disk seems unlikely to reproduce the evolution from Class\,I to Class\,II that we observe. But it is interesting to notice that the slope of the MHD wind model can reproduce the trend of the only Class\,I sample if shifted toward the top of the plot, i.e. by assuming higher $\macc$. 
However, the plot qualitatively suggests that the hybrid model can better represents both the Class\,I ans II data distribution, therefore some wind contribution should be included in the viscous evolution to best reproduce the observations. 
\\ \\
Limitations of comparing our observational results with models lies in the following two reasons. 
First, these models do not investigate the earliest stages of the star formation, in other words, both models by \citet[][]{lodato17} and \citet[][]{Tabone.Rosotti.ea2021} set the disk mass to a fixed value which dissipates with time as mass is accreted on the forming star, while in the protostellar phase the disk mass is replenished by the envelope. 
Second, both the magnetic effects and the disk viscous timescale depends on the environmental effects, and our sample is composed by Class\,I stars belonging to very different regions in the solar neightbothood. 
Naturally, observations of single star-forming clouds and models that include the earliest stages of disk formation are necessary to further constrain the disk wind and viscous models. 
Also, the disk masses are highly uncertain, so obtaining disk masses at longer wavelengths like the upcoming Band\,1 of ALMA or the shortest VLA wavelengths would improve this analysis.
Moreover, the uncertainty on the protostars' age propagates to the mass accretion rate, providing uncertainties larger than for Class\,II PMS stars.

\subsection{$\macc \, vs. \, \mdisk$ relation and planets formation}
Assuming that planets form by accreting materials onto planetesimals, and given some assumptions on the disk structure and evolution, in the last decade many population synthesis models were developed to describe different kind of produced exoplanetary systems \citep[see][for a review]{benz2014}. 
In particular, the population synthesis of planet formation by \citet{lubow.dangelo2006} predicts a population of disks with greatly decreased accretion rates onto protostars due to the presence of gas giant planets. 
We do not see this population in the $\dot{M}_{\rm acc}$\,--\,$M_{\rm disk}$ plot, as \citet{man19} for CTTSs, showing that this effect is not present even in younger disks. 
It is unclear if those disks are massive enough to host gas giants or perhaps this effect is not present in general. 

\section{Summary and Conclusions}
We presented for the first time in this letter  the $\macc \, vs. \, \mdisk$ plot populated with also Class\,I YSOs, shifting to the protostellar stage the investigation of the disk initial conditions, crucial to understand the star and planet formation.
Our data show that younger sources present higher mass accretion rate, more massive disks in general, and have ``depletion times" (i.e. $\mdisk/\macc$) faster than Class\,II YSOs suggesting an evolutionary trend between Class\,I and II YSOs. We also measure higher $\macc$ in Class\,I than in Class\,II with the same disk mass. 
Since our sample is limited to brightest and, thus, older sources among Class\,I, we can consider our results as lower limits for Class\,I in general.

We fitted the $\macc/\mdisk$ distribution of our Class\,I sample finding a slope flatter than the corresponding slope for Class\,II sources. But focusing only on the more massive disks, we found the Class\,I slope is in agreement with the Class\,II slope, suggesting that differences between the $\macc - \mdisk$ distributions of Class\,I and II should be investigated in protostars whose disk mass is comparable with typical Class\,II $\mdisk$.

We tested our results with most recent viscous and MHD wind models. We tentatively speculate that our data can be described by viscous model together with some contamination by MHD winds (hybrid model). 
However, we find no definitive conclusions about which of these models better represent our data. We associated this to the absence of an envelope feeding the disk in these models, and to the fact that even if our sample is analysed in an homogeneous way, it is affected by different environmental effects, since these protostars belong to different star-forming regions.

Uniform samples of Class\,I and Class\,II protostars with identical initial conditions, i.e. in the same star-forming region, and theoretical models which describe both stages are necessary to draw solid conclusions on the evolutionary path of YSOs and to be able to set the initial conditions for stars and planets formations. 
While VLT/KMOS can be used efficiently on larger samples, JWST will deliver most sensitive information on photospheres with NIRSpec and eventually enable the investigation of the protostellar accretion rates for even more embedded sources with MIRI.

\begin{acknowledgments}

We thank the anonymous referee for the useful comments which improve this manuscript.
This project has received funding from the European Research Council (ERC) under the European Union's Horizon 2020 Research \& Innovation Programme under grant agreement No 716155 (SACCRED), and by the European Union under the European Union’s Horizon Europe Research \& Innovation Programme 101039452 (WANDA). Views and opinions expressed are however those of the author(s) only and do not necessarily reflect those of the European Union or the European Research Council. Neither the European Union nor the granting authority can be held responsible for them.
GR acknowledges support from the Netherlands Organisation for Scientific Research (NWO, program number 016.Veni.192.233) and from an STFC Ernest Rutherford Fellowship (grant number ST/T003855/1).

\end{acknowledgments}



\software{astropy \citep{2013A&A...558A..33A,2018AJ....156..123A}, astroquery, matplotlib.
          }

\appendix

\section{The Sample}
\label{app:sample}
Tab.\,\ref{tab:first_res} lists the main properties of the protostars included in the analysis. 

\startlongtable
\begin{deluxetable*}{lccccc}
\tablecaption{\label{tab:first_res}}
\tablewidth{0pt}
\tablehead{
\colhead{Name} &	Cloud & Distance & $\macc$ & $M_{dust}$ & ref.\\
\colhead{  }  &       & pc        & $\msun$/yr & $\mearth$}
\decimalcolnumbers
\startdata
IRS2	                & CrA           & $160.5\pm1.8 $  $^a$ &  $3\times 10^{-7}$	        & $693.8\pm1.6$ &  1, 4    \\
IRS5a	                & CrA           & $160.5\pm1.8 $  $^a$ & $3\times 10^{-8}$          & $587.1\pm2.1$  & 1, 4   \\
HH 100 IR	                & CrA           & $160.5\pm1.8 $  $^a$ &  $2\times 10^{-6}$        & $443.5\pm8.0$ & 1, 4\\
HH 26 IRS               & L1630	        & 450                  & $8.5 \times 10^{-7}$       & $468.7\pm18.1 $ & 2, 5 \\
HH 34 IRS               & L1641         &	460                & $41.1 \times 10^{-7}$      & $ 2282.7\pm67.7 $ & 2, 5 \\
HH 46 IRS               & Bok globule   & 450                  & $2.2 \times 10^{-7}$       &$ 1126.8\pm14.5 $ & 2, 6 \\
2MASSJ03283968+3117321   & NGC\,1333     & $293 \pm 22$ $^b$        & $(1.4-8.3) \times 10^{-8}$ & $12.8\pm0.8$ & 3, 6 \\ 
2MASSJ03285842+3122175   & NGC\,1333     & $293 \pm 22$ $^b$        & $(19-70  ) \times 10^{-8}$ & $11.3\pm0.6$ & 3, 6\\ 
2MASSJ03290149+3120208   & NGC\,1333     & $293 \pm 22$ $^b$        & $(16-38  ) \times 10^{-8}$ & $7.7\pm2.2$  & 3, 6\\ 
SVS 13 (V512 Per)$^{\dagger}$ & NGC\,1333 & $293 \pm 22$ $^b$        & $(19-220 ) \times 10^{-8}$ &  $969.7\pm15.5$  & 3, 7 \\ 
LAL96 213               & NGC\,1333     & $293 \pm 22$  $^b$       & $(13-120 ) \times 10^{-8}$ & $318.5\pm0.9$ & 3, 6 \\ 
2MASSJ03292003+3124076   & NGC\,1333     & $293 \pm 22$ $^b$        & $(0.8-3.6) \times 10^{-8}$ & $3.8 \pm1.6$ & 3, 8 \\ 
CG2010IRAS032203035N 	& Per-IC348 	& $219.8\pm16.2 $  $^{a}$ & $(3.2 - 9.3) \times 10^{-8}$ & $167 \pm  25$ & this work, 7\\			
2MASSJ033312843121241 	& Per-IC348 	& $319.5\pm23.7	$  $^{a}$ & $(1.5 - 3.0) \times 10^{-7}$ & $190 \pm 28$ & this work, 6\\           
BHS98MHO1 				& Tau-L1495 	& $134.0\pm7.0 $   $^{a}$ & $(1.2 - 3.3) \times 10^{-7}$ & $171 \pm 18$ & this work, 9\\           
BHS98MHO2 				& Tau-L1495 	& $131.0\pm2.9$    $^{a}$ & $(2.3 - 5.9) \times 10^{-8}$ & $100.5 \pm 4.5$ & this work, 9\\        
IRAS041692702 			& Tau-L1495 	& $129.5 \pm 12.9 $ $^c$  & $(3.6 - 8.3) \times 10^{-8}$& $147 \pm   47$ & this work, 10\\         
VFSTau 					& Tau-Aur 		& $133.9\pm2.4$ $^{a}$    & $(1.2 - 3.2) \times 10^{-7}$ & $1.79 \pm  0.13$ & this work, 11\\ 
2MASSJ042200692657324 	& Tau-Aur 		& $133.9\pm2.4$ $^d$ 	  & $(3.9 - 4.8) \times 10^{-7}$ & $139  \pm  15$ & this work, 12\\        
IRAS042952251 			& Tau-L1546 	& $160.76 \pm 16.1$ $^c$  & $(2.1 - 4.4) \times 10^{-8}$ & $ 125 \pm 62$ & this work, 10\\         
IRAS043812540 			& Tau-L1527 	& $141.8 \pm 1.4$	$^c$  & $(3.2 - 5.9) \times 10^{-8}$ & $24.6 \pm 4.9$ & this work, 13\\        
Parenago2649 			& ONC A 		& $398.5\pm2.5 $	 $^a$ & $(0.4 - 3.3) \times 10^{-7}$ & $ 105 \pm 21$ & this work, 14\\         
2MASSJ054050590805487 	& ONC A 		& $440 \pm 44 $	$^e$	  & $(3.6 - 8.1) \times 10^{-8}$ & $72 \pm 15$ & this work, 15\\           
2MASSJ054049910806084 	& ONC A 		& $440 \pm 44 $	$^e$	  & $(1.3 - 3.7) \times 10^{-7}$ & $10.2 \pm 2.7$ & this work, 15\\        
IRAS054050117 			& ONC B 		& $420 \pm 42 $	$^e$	  & $(1.5 - 4.2) \times 10^{-7}$ & $ 191 \pm 38$ & this work, 15\\         
VSCrA 					& CrA 			& $160.5\pm1.8 $  $^a$    & $(0.5 - 2.1) \times 10^{-8}$  & $306.7 \pm  6.9 $ & this work, 4
\enddata
\tablecomments{$^a$Parallax distance with {\it Gaia} EDR3 direct match \citep{GaiaCollaboration.Brown.ea2021}, Distance to the region (error is set to 10\% if not stated in literature): $^b$\cite{ort18}, $^c$\cite{Krolikowski.Kraus.ea2021},   
$^d$assumed to be the same as FS TauA where {\it Gaia} EDR3 is available, $^e$\cite{tob20}. 
References: 1\,-\,\cite{nis05a}, 2\,-\,\cite{ant08}, 3\,-\,\cite{fio21}, 4\,-\,ALMA\#2019.1.01792.S, 5\,-\,\cite{tob20}, \cite{Zhang.Arce.ea2016},  6\,-\,\cite{tyc20}, 7\,-\,\cite{Tobin.Looney.ea2018}, 8\,-\,\cite{Yang.Sakai.ea2021}, 9\,-\,\cite{Akeson.Jensen2014}, 10\,-\,\cite{Sheehan.Eisner2017}, 11\,-\,\cite{Akeson.Jensen.ea2019}, 12\,-\,\cite{Villenave.Menard.ea2020}, 13\,-\,\cite{vantHoff.Harsono.ea2020}, 14\,-\,ALMA\#2019.1.01813.S, 15\,-\,\cite{tob20}. Mass accretion rates correspond to the range of values inferred assuming the age varying from the birthline to 1\,Myr.}
\end{deluxetable*}


\bibliography{sample631}{}
\bibliographystyle{aasjournal}



\end{document}